# Band Structure Engineering of 2D materials using Patterned Dielectric Superlattices

Carlos Forsythe[1], Xiaodong Zhou[1,2], Takashi Taniguchi[3], Kenji Watanabe[3], Abhay Pasupathy[1], Pilkyung Moon[4], Mikito Koshino[5], Philip Kim[6], Cory R. Dean[1]

[1]Department of Physics, Columbia University, New York, New York 10027, USA
[2]Laboratory of Advanced Materials, Fudan University, Shanghai 200438, P. R. China
[3]National Institute for Materials Science, 1-1 Namiki, Tsukuba 305-0044, Japan
[4] NYU Shanghai, Shanghai 200122, China
  NYU-ECNU Institute of Physics at NYU Shanghai, Shanghai 200062, China
[5]Department of Physics, Osaka University, Toyonaka, 560-0043, Japan
[6]Department of Physics, Harvard University, Cambridge, Massachusetts 02138, USA

**The ability to manipulate two-dimensional (2D) electrons with external electric fields provides a route to synthetic band engineering. By imposing artificially designed and spatially periodic superlattice (SL) potentials, 2D electronic properties can be further engineered beyond the constraints of naturally occurring atomic crystals**[1–5]**. Here we report a new approach to fabricate high mobility SL devices by integrating surface dielectric patterning with atomically thin van der Waals materials. By separating the device assembly and SL fabrication processes, we address the intractable tradeoff between device processing and mobility degradation that constrains SL engineering in conventional systems. The improved electrostatics of atomically thin materials moreover allows smaller wavelength SL patterns than previously achieved. Replica Dirac cones in ballistic graphene devices with sub 40nm wavelength SLs are demonstrated, while under large magnetic fields we report the fractal Hofstadter spectra**[6–8] **from SLs with designed lattice symmetries vastly different from that of the host crystal. Our results establish a robust and versatile technique for band structure engineering of graphene and related van der Waals materials with dynamic tunability.**

Electrons in a periodic lattice, such as in an atomic crystal, develop a band structure whose characteristics are determined by the spacing, symmetry, and magnitude of the lattice site potentials. Since in 2D, electrons can additionally be manipulated by externally applied electrostatic fields, application of spatially periodic potentials can be utilized to modify the intrinsic material band structures[1–5]. Atomically thin 2D materials, such as graphene and the more recently studied transition metal dichalcogenides[9], provide promising new platforms to realize synthetic band structures by SL pattering. In graphene, novel effects due to the linear band dispersion are predicted, such as anisotropic band renormalization and electron collimation at low density and the formation of replica Dirac cones and tunable Van Hove singularities at large density[10–12]. Moreover, monolayer atomic crystals such as graphene are directly accessible, providing the possibility for much smaller SL wavelengths than in conventional 2D heterostructures where the active layer is generally buried 30 nm or more[2–5].

SL engineering in graphene has been pursued extensively utilizing electrostatic gating schemes[13–15] as well as direct modification by chemical doping[16], etching [17–19], and strain engineering[20]. In all of these efforts the primary technical challenge remains the SL length scale. For SL's that are too large, the associated SL mini-Brillouin zone edge, which scales inversely with the SL lattice constant, appears too close to the neutrality point where it becomes obscured by disorder. On the other hand, too small a SL lattice constant causes the zone edge energy to become so large that it is not accessible by conventional gating schemes. This results in a Goldilocks scenario requiring the SL lattice constant to be of order a few tens of nanometers to be observable. Reaching this regime in any material system has been challenging due to limited resolution achievable by lithographic patterning, and the intractable tradeoff between device processing and electron mobility[4,5,13,14]. In graphene on hexagonal-boron nitride (hBN) a moiré pattern, resulting from the slightly mismatched lattice constants, provides a naturally occurring SL potential with a nearly ideal length scale of ∼14 nm while preserving high mobility[21]. Indeed, the superior characteristics of moiré-SLs have enabled several remarkable observations including modified electron dynamics[22], emergent band-gaps[8] and demonstration of the fractal Hofstadter "Butterfly" band structure when combined with magnetic fields[6,7]. However, moiré-SLs suffer from lack of tunability in either the SL symmetry or magnitude, limiting their application for engineering electronic structures.

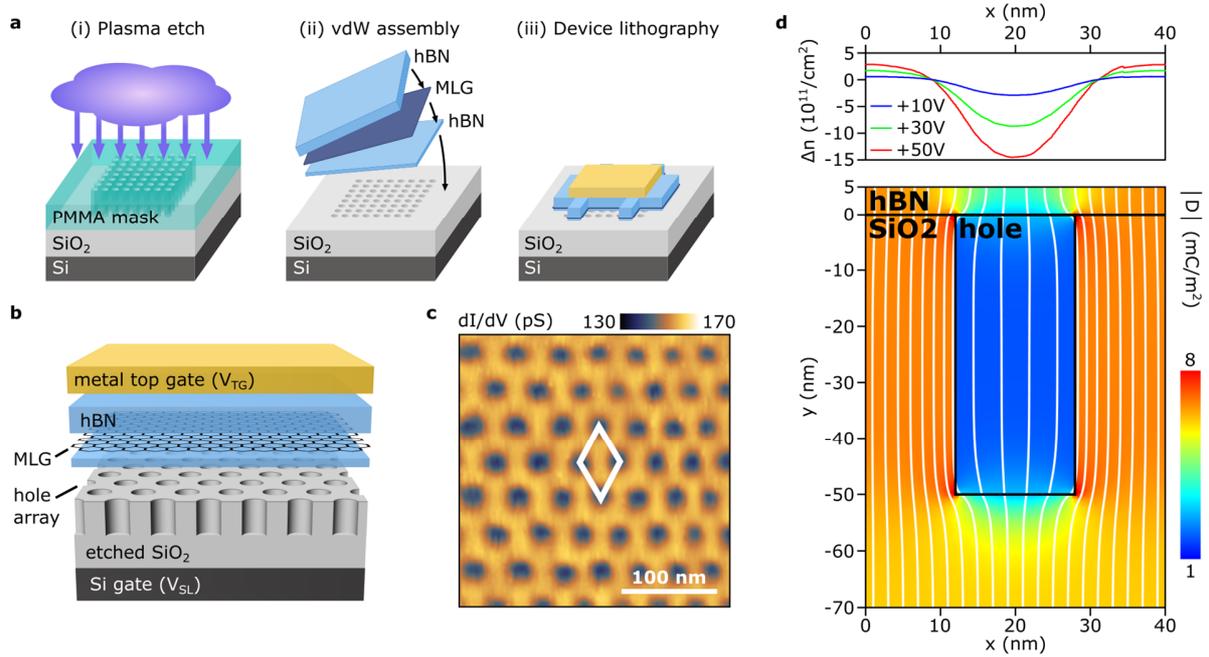

**Figure 1, dielectric superlattice design. a**, A schematic diagram shows the fabrication process used to impose an electrostatically defined superlattice potential onto monolayer graphene. Using standard lithographic processes, we first etch a periodic array of holes into a SiO$_2$ surface that is thermally grown on a highly doped Si wafer (see Methods for more details). Next we mechanically assemble and transfer an hBN-encapsulated graphene structure onto the hole array. Finally, we shape the device into a Hall bar and deposit edge contacts, using previously described techniques[23]. **b**, Cartoon schematic of a final device. **c**, Variation of the density of states ($DOS$) imaged directly through STM dI/dV measurement of a device with exposed graphene under $V_{SL} = -25$ V. **d**, A model of the displacement field below the graphene layer in a single cell of the PDSL including field lines. Also plotted is the relationship between the applied Si gate voltage and the resulting density variation, $\Delta n$, induced in the graphene channel, using typical device parameters (see SI).

Here we report a novel approach to realize electrostatically defined graphene SLs. In conventional electrostatic-based schemes for semiconductor heterostructures, the SL potential is typically achieved by either introducing a spatially patterned gate electrode[2,4,5], or patterning the ionic donor layer[24]. In both cases, remote electrostatic modulations are transferred capacitively to the 2D electron gas (2DEG). In our new approach, we have developed an inverted scheme where we utilize a uniform, featureless gate electrode and instead modulate the dielectric interface in the local vicinity of the 2DEG. The patterned dielectric superlattice (PDSL) scheme provides two critical advantages. First, electrostatic boundaries can be defined as sharply as the encapsulating dielectric, which, in the case of hBN, can be made as thin as 1nm[25,26]. Second, using the van der Waals (vdW) transfer technique[23], we separate the PDSL and device fabrication processes, allowing us to independently optimize each before bringing them together (Fig. 1a). Using these techniques, we fabricate PDSLs with periodicities as low as 35nm, while maintaining ultra-high mobilities typical of BN encapsulated graphene. We observe SL Dirac peaks for both square and triangular lattice symmetries, and under finite magnetic field demonstrate the capability to realize fully developed fractal mini-gaps of the Hofstadter spectrum for non-moiré patterned SLs. Our results establish PDSL-vdW heterostructures as a versatile new platform for synthetic band structure engineering.

Fig. 1b shows a cartoon schematic of a typical graphene device geometry. Applying a bias to the bottom doped Si gate results in a modulated potential in the graphene layer. In order to maximize device mobility we encapsulate the graphene with boron nitride using the dry transfer technique[23], but we restrict the bottom BN layer thickness to the few nm limit in order to maximize the modulation amplitude. A second top gate electrode is used to independently tune the carrier density in the channel. Using a simple capacitor model we calculate the average carrier density according to, $n = (C_{TG}/e)V_{TG} + (C_{SL}/e)V_{SL} + n_0$, where $V_{TG}$ and $V_{SL}$ are the biases applied to the top metal gate and bottom Si gate electrodes, respectively. The geometric capacitances, $C_{TG}$ and $C_{SL}$, as well as the intrinsic doping, $n_0$, are measured from zero field transport and the Hall response of the device. The SL density modulation in the channel is determined by the SL gate bias, $V_{SL}$, with the precise magnitude and profile dependent on the shape of the SL and dielectric properties. An example of a calculated density profile is shown in Fig. 1d for typical device parameters, obtained using a commercially available electrostatic modeling package (see SI). Fig. 1c shows a scanning tunneling microscopy (STM) image acquired from a test device with a 12 nm bottom

BN layer, but no top BN layer and a triangular PDSL with wavelength $\lambda = 40$ nm. The device exhibits a spatially periodic variation in the density of states ($DOS$) of excellent uniformity due to the PDSL induced carrier density modulation. Further STM measurements comparing PDSL site and anti-site Dirac point energies show reasonable agreement with our electrostatic model (see SI).

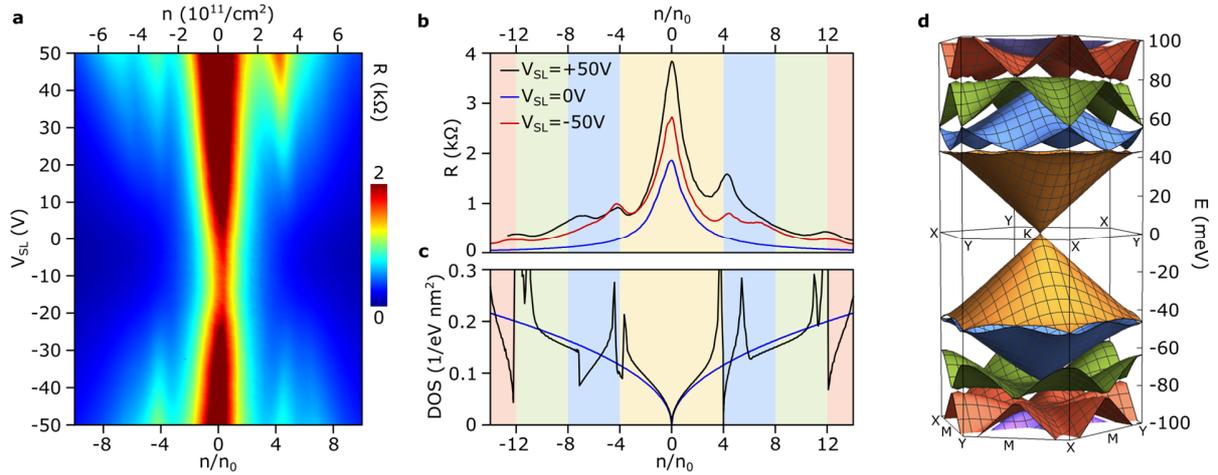

**Figure 2, Transport data vs. theory for a triangular superlattice device. a**, Device resistance as a function of normalized carrier density, $n/n_0$, and superlattice voltage, $V_{SL}$ measured at $T = 250$ mK. $n_0$ is the superlattice cell density (see text). At $V_{SL} = 0$ V, there is a single maximum in resistance at $n = 0 \times 10^{11}$ cm$^{-2}$, matching expectation for a normal graphene device. As $|V_{SL}|$ increases, the SL modulation strengthens and new features in resistance develop. **b,** Line cuts of device resistance at $V_{SL} = +50$ V, $0$ V, $-50$ V. Under $V_{SL} = 0$ V, the simple graphene behavior is reclaimed. Device resistance at $V_{SL} = +50$ V and $V_{SL} = -50$ V show the system under maximum SL strength, but opposite relative polarity **c,** The modeled $DOS$ and **d,** band structure of a graphene system under an additional superlattice potential approximating our system at $V_{SL} = +50$ V. Minima in the DOS align well with experimental resistance maxima for the trace at $V_{SL} = +50$ V.

Figure 2a shows transport measurements from a triangular PDSL device with wavelength, $\lambda = 40$ nm. We plot the measured resistance as a function of SL gate voltage, $V_{SL}$, and carrier density, $n$, measured at $T = 250$ mK. For $V_{SL} = 0$ V, there is only a single peak in resistance at the charge neutrality point (CNP), as expected for intrinsic graphene. We measure a mobility of 300,000 cm$^2$/Vs at high density, corresponding to a mean free path in excess of the device's 4 μm device width, indicating ballistic transport across the device, consistent with expectation for an hBN encapsulated graphene system 2DEG[23]. This suggests that the patterned SiO$_2$ substrate does not significantly degrade the device performance.

As $V_{SL}$ increases, the SL potential increases in strength, resulting in new resistive peaks, symmetrically located around the CNP. Single line cuts along varying density, corresponding to a fixed bias on the SL gate are shown in Fig. 2b. Plotting the resistance against normalized density, $n/n_0$, where $n_0 = 1/A$ and $A$ is the area of the superlattice unit cell, the strongest satellite resistance peaks appear to coincide with $n/n_0 = \pm 4$. These densities correspond with filling to the SL cell Brillouin zone boundary where it is expected that newly generated massless Dirac fermions are formed[27]. The factor of 4 is a result of the spin/valley degeneracy in graphene. The band structure and $DOS$, calculated by including potential terms in the Hamiltonian to approximate the 40nm triangular PDSL with SL gate voltages $V_{SL} = +50$ V, are shown in Figs. 2d,c (see SI). There is excellent agreement between the theoretical $DOS$ and measured resistance for the same SL voltage. In particular, we find that peaks in the resistance correlate well with regions of expected $DOS$ minimization. We note that the $DOS$ minimization at $n/n_0 = +4$ is much stronger than that at $n/n_0 = -4$, owing to electron-hole asymmetry in the band structure, including a slight band overlap in the hole region, compared to well defined linear crossings in the electron region (Fig. 2d). This is consistent with a clear asymmetry in the transport response where the electron side satellite peak is more resistive than the hole side. Upon reversing the sign of the SL bias to $V_{SL} = -50$ V, we again observe all of the same features, however the asymmetry is reversed, consistent with an inverted SL potential. We note that satellite Dirac peaks (SDPs) near $n/n_0 = \pm 4$ are characteristic features of the SL previously identified in moiré patterned graphene/BN heterostructures, but the density where the SDPs occur in the moiré-SLs is larger by an order of magnitude due to their much shorter maximal periodicity

($\sim$14 nm)[21]. A similar response was measured for a square dielectric SL with a periodicity of $\lambda = 35$ nm, displaying features in excellent agreement with the calculated band structure for a square PDSL (see SI).

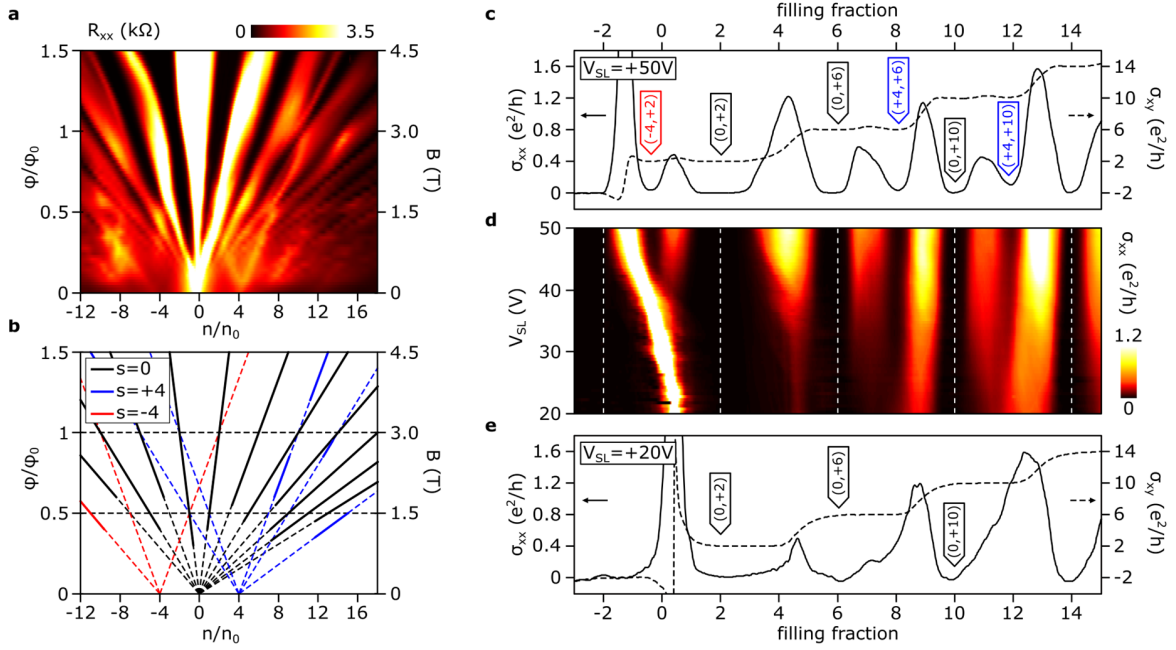

**Fig. 3. Hofstadter spectrum in magnetic field. a,** Device resistance at $V_{SL} = +50$ V and $T = 250$ mK for the triangular superlattice device, as a function of normalized density, $n/n_0$, and normalized magnetic field $\phi/\phi_0$ (see text). The SL potential leads to a clear satellite fan centered at $n/n_0 = +4$. The maximum plotted field of $\phi/\phi_0 = 1.5$ corresponds to $B = 4.5$ T. **b,** Wannier diagram highlighting traces of resistance minima from the plot **a**. The Hofstadter mini-gaps gaps associated with the replicated Dirac points (i.e., $s \neq 0$), are colored red ($s = -4$) and blue ($s = 4$) respectively. **c-e,** longitudinal conductance, $\sigma_{xx}$ (solid lines), and transverse conductance, $\sigma_{xy}$ (dashed lines), of the triangular device at $B = 5.5$ T ($\phi/\phi_0 = 1.8$) as functions of Landau level filling fraction, at superlattice gate voltage **c,** $V_{SL} = +50$ V and **e,** $V_{SL} = +20$ V. Plateaus in the Hall conductance are labeled with by their $s$ and $t$ quantum numbers (see text). Features unique to the Hofstadter spectrum ($s \neq 0$), require a sufficiently large superlattice gate voltage, $V_{SL}$, to manifest. Minima in $\sigma_{xx}$ and plateaus in $\sigma_{xy}$ associated with these states are distinct from their neighboring conventional quantum Hall features at $V_{SL} = +50$ V. **d,** $\sigma_{xx}$ as a function of filling

fraction and $V_{SL}$. Hofstadter mini-gaps disappear as $V_{SL}$ is lowered continuously from $V_{SL} = +50$ V to $+20$ V.

To further characterize the effect of the SL potential we examined magnetoresistance in the quantum Hall effect (QHE) regime. A hallmark of SL systems is the emergence of a fractal "butterfly" spectrum under strong magnetic fields, a consequence of the interplay between the electrostatic SL and magnetic length scales[28]. Fig. 3a shows the longitudinal resistance, $R_{xx}$, as a function of carrier density and applied magnetic field for the same triangular lattice device as in Fig. 2. The magnetic field axis is expressed in units of $\phi/\phi_0$, where $\phi$ is the flux per SL unit cell and $\phi_0$ is the quantum of magnetic flux. In addition to the set of conventional Landau gaps, Hofstadter mini-gaps are also well resolved. Moreover, several features follow a fan-like shape converging to $n/n_0 = +4$ consistent with the zero field peak corresponding to the SL Brillouin zone boundary. The QHE features of the Hofstadter's fractal spectrum can be described by two topological integers, $(s,t)$, satisfying a Diophantine equation: $\frac{n}{n_0} = t\frac{\phi}{\phi_0} + s$. A Wannier diagram identifying the observed Hofstadter mini-gaps is shown in Fig. 3b[8]. Due to the long wavelength of the PDSL, the magnetic field at which $\phi/\phi_0 = 1$, corresponding to the condition for precisely one flux quanta per SL unit cell in the system, can be reached at only $B = 3.0$ T. This compares favorably to previously studied moiré-SLs where magnetic fields in excess of 20T are necessary[6–8,21].

In figure 3c-e we plot transport data under a constant magnetic field ($B = 5.5$ T, $\phi/\phi_0 = 1.8$) but with varying bias on the SL gate. At large $V_{SL}$, we observe a sequence of well-developed Hofstadter mini-gap states, namely zero valued longitudinal conductivity simultaneous with well quantized hall conductivity, with non-zero $s$ number (labeled in the single line trace in Fig. 3c). As the SL gate voltage is continuously lowered from $V_{SL} = +50$ V to $V_{SL} = +20$ V, this external potential abates and mini-gap states disappear, leaving only the conventional Landau gaps (Fig. 3 d,e). This result confirms that our device architecture provides the capability to dynamically tune the band structure by simply varying the SL gate bias.

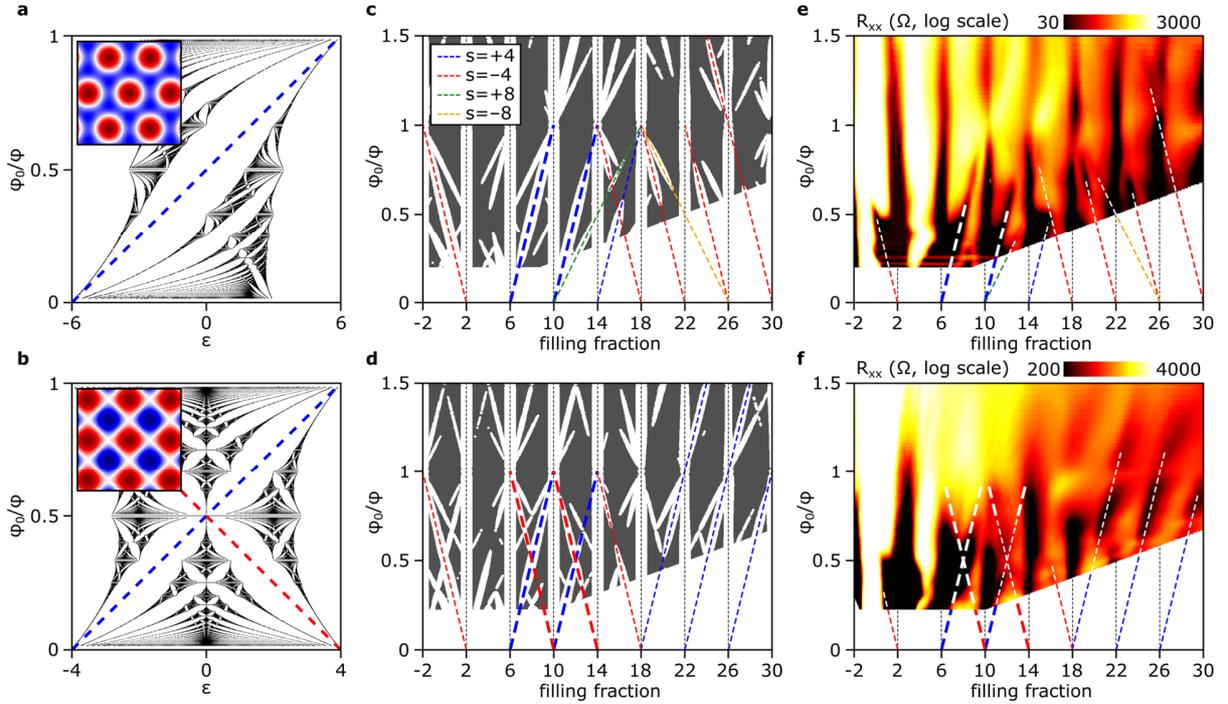

**Fig. 4. Magnetic sub-band structure of triangular and square superlattices. a,b** Theoretical "Unit Cells" depicting the lowest Landau level sub-bands of a system under both a quantizing magnetic field and **a,** a triangular and **b,** a square lattice potential perturbation. Spectra are plotted against a unitless scaled energy[29], $\varepsilon$, and $\phi_0/\phi \propto 1/B$. Insets: plots of lattice potentials used in the derivation of these unit cells. Between the positive (red) and negative (blue) regions of these potentials, the triangular lattice is asymmetric while the square is symmetric under reversing the sign of the potential. **c,d** Wannier plot of the theoretically calculated gaps for triangular and square lattice, respectively (see SI). **e,f** Longitudinal resistance, $R_{xx}$, of triangular and square superlattice devices, respectively, plotted against inverse normalized flux, $\phi_0/\phi$, and filling fraction, with the same parameters as modelled in c,d. The resistance mimina closely follow the position of expected mini-gaps. Triangular and square lattices display different gap structures owing to differences in the lattice symmetry.

Unlike moiré-SLs, which follow the hexagonal symmetry of underlying graphene lattice, the symmetry of PDSLs can be designed arbitrarily. Fig. 4 shows magnetotransport response for a triangular and square PDSL for comparison. The calculated energy spectra, shown in Figure 4a,b, exhibit qualitative differences. Most notably, while the fractal spectrum for a square superlattice is highly symmetric, the triangular superlattice spectrum is dominated by a single diagonal gap. These relative trends are expected to be evident in the magnetotransport at high field as observed in the theoretically calculated Wannier diagrams shown in Fig. 4c,d, where we have plotted the largest theoretically calculated gaps. We note that owing to a small SL strength (~50meV at the highest applied SL gate bias at relevant densities, see SI) relative to the Landau gaps, we assume the SL to be in the weak coupling limit[30,31]. In this limit, it is more convenient to express the Hofstadter's mini gaps using inverse normalized flux, $\phi_0/\phi$ and Landau level filling fraction, $v = (n/n_0)(\phi_0/\phi)$[28–31]. Plotted on these axis, gap trajectories in the Wannier diagram are described by an equivalent Diophantine relation $v = s \cdot \frac{\phi_0}{\phi} + t$ [31,32]. In Fig. 4c,d, the calculated gaps are shown as white lines with thickness proportional to the gap size.

Our experimental data (Fig. 4e,f) show excellent correspondence to the calculated spectra. The asymmetric nature of the triangular unit cell is identifiable in the 2$^{nd}$ and 3$^{rd}$ Landau bands ($v = 6 - 10$ and $10 - 14$ respectively), where there are single strong gaps with $s = +4$ (Fig. 4e) dominating these sub-bands, matching the dominant diagonal gap of the triangular unit cell. In the same region of square lattice magnetotransport (Fig. 4f), traces of both $s = +4$ and $s = -4$ gaps emerge, consistent with the more symmetric crossing of the square unit cell spectrum. The special symmetry of the square lattice is related to the sign reversal symmetry of the sinusoidal SL potential (Fig. 4b inset). While these symmetry arguments apply specifically to the lowest Fourier components of triangular and square lattices (depicted in Fig 4a,b insets), these long wavelength components have an exponentially stronger impact on the fractal spectrum than higher order components also present in PDSLs[29]. We also note that due to the relatively long SL periodicity in these devices, we can reach values of $\phi_0/\phi$ as low as $\phi_0/\phi = 0.2$ at fields of only 15 Tesla; a condition which would require fields over 100T in moiré-SL systems.

We have demonstrated a novel approach to fabricating a nanoscale electrostatic superlattice onto a high mobility graphene system by engineering the dielectric environment of the 2DEG rather than the doping layer. We demonstrate strong modification of the graphene band structure without substantial mobility degradation. Furthermore, we demonstrate the first full characterization of a square lattice Hofstadter spectrum at large magnetic fields (Fig 4f) and its comparison to a triangular lattice system finding excellent qualitative agreement with their respective band structures. Lithographically defined PDSLs can be applied to a wide array of SL symmetries, and can be integrated with the broader class of 2D van der Waals materials[33].

**METHODS:** Our PDSLs were fabricated through plasma etching of $SiO_2$ using a thin PMMA mask. We used 495 A2 PMMA spun to 50 nm thickness onto a doped Si wafer with thermally grown $SiO_2$ of thickness ~300 nm. Lattices were written as an array of circles at $i = 100 - 400$ pA current using different e-beam tools: the Nanobeam nB4 at Columbia University and an Elionix ELS-G100 at the ASRC Nanofabrication Facility. Holes were plasma etched using an Oxford Plasmalab 80 Plus system with a mixture of $CHF_3$ and Ar gas, at flow rates of 40sccm and 5sccm respectively, to a depth of roughly 50nm. The PDSLs were then cleaned using $O_2$ plasma followed by piranha etching. Graphene-hBN stacks were transferred onto the PDSLs using a co-lamination mechanical transfer technique[23]. Bottom hBN thicknesses of 5nm and 3nm were used for the triangular and square lattice devices, respectively. These stacks were fabricated into Hall bar devices using standard e-beam lithography processes. Our devices have a channel width of 4um and spacing between voltage probes of 5.5um and 6um for the triangular and square lattice devices, respectively. Transport measurements were taken in a 15 Tesla Oxford $^3$He system (sample in vapor) at temperature, $T \cong 250$mK. Voltage measurements were acquired using SR830 lock-in amplifiers at a current bias of 100nA.

**ACKNOWLEDGEMENTS**: This work was supported by the National Science Foundation (DMR-1462383). C.F. was supported by NSF GRFP (DGE-14-44869) and NSF IGERT (DGE-1069240). This work was performed in part at the Advanced Science Research Center NanoFabrication Facility of the Graduate Center at the City University of New York. A portion of this work was performed at the National High Magnetic Field Laboratory, which is supported by National Science Foundation Cooperative Agreement No. DMR-0654118, the State of Florida and the U.S. Department of Energy. P.M. acknowledges the support of NYU Shanghai (Start-up Funds), NYU-ECNU Institute of Physics, and the NSF of China Research Fund for International Young Scientists (Grant No. 11550110177). This research was carried out on the High Performance Computing resources at NYU Shanghai. MK was supported by JSPS KAKENHI (Grant No. JP25107005, JP25107001 and JP15K21722). PK acknowledges support from ONR MURI on Quantum Optomechanics (Grant No. N00014-15-1-2761).

1. Esaki, L. & Tsu, R. Superlattice and Negative Differential Conductivity in Semiconductors. *IBM J. Res. Dev.* **14,** 61–65 (1970).

2. Ismail, K., Chu, W., Yen, A., Antoniadis, D. A. & Smith, H. I. Negative transconductance and negative differential resistance in a grid-gate modulation-doped field-effect transistor. *Appl. Phys. Lett.* **54,** 460–462 (1989).

3. Weiss, D., Klitzing, K. V., Ploog, K. & Weimann, G. Magnetoresistance Oscillations in a Two-Dimensional Electron Gas Induced by a Submicrometer Periodic Potential. *EPL* **8,** 179 (1989).

4. Schlösser, T., Ensslin, K., Kotthaus, J. P. & Holland, M. Landau subbands generated by a lateral electrostatic superlattice - chasing the Hofstadter butterfly. *Semicond. Sci. Technol.* **11,** 1582–1585 (1996).

5. Albrecht, C. *et al.* Evidence of Hofstadter's Fractal Energy Spectrum in the Quantized Hall Conductance. *Phys. Rev. Lett.* **86,** 147–150 (2001).

6. Dean, C. R. *et al.* Hofstadter's butterfly and the fractal quantum Hall effect in moire superlattices. *Nature* **497,** 598–602 (2013).

7.  Ponomarenko, L. A. *et al.* Cloning of Dirac fermions in graphene superlattices. *Nature* **497,** 594–597 (2013).

8.  Hunt, B. *et al.* Massive Dirac Fermions and Hofstadter Butterfly in a van der Waals Heterostructure. *Science* **340,** 1427–1430 (2013).

9.  Hua Wang, Q., Kalantar-Zadeh, K., Kis, A., Coleman, J. N. & Strano, M. S. Electronics and optoelectronics of two-dimensional transition metal dichalcogenides. *Nat. Nanotechnol.* **7,** 699–712 (2012).

10. Park, C.-H., Yang, L., Son, Y.-W., Cohen, M. L. & Louie, S. G. Anisotropic behaviours of massless Dirac fermions in graphene under periodic potentials. *Nat. Phys.* **4,** 213–217 (2008).

11. Park, C. H., Son, Y. W., Yang, L., Cohen, M. L. & Louie, S. G. Electron beam supercollimation in graphene superlattices. *Nano Lett.* **8,** 2920–2924 (2008).

12. Barbier, M., Vasilopoulos, P. & Peeters, F. M. Extra Dirac points in the energy spectrum for superlattices on single-layer graphene. *Phys. Rev. B* **81,** 75438 (2010).

13. Dubey, S. *et al.* Tunable Superlattice in Graphene To Control the Number of Dirac Points. *Nano Lett.* **13,** 3990–3995 (2013).

14. Drienovsky, M. *et al.* Towards superlattices: Lateral bipolar multibarriers in graphene. *Phys. Rev. B* **89,** 115421 (2014).

15. Drienovsky, M. *et al.* Few-layer graphene patterned bottom gates for van der Waals heterostructures. *arXiv* **1703.05631,** (2017).

16. Sun, Z. *et al.* Towards hybrid superlattices in graphene. *Nat. Commun.* **2,** 559 (2011).

17. Bai, J., Zhong, X., Jiang, S., Huang, Y. & Duan, X. Graphene nanomesh. *Nat. Nanotechnol.* **5,** 190–194 (2010).

18. Sandner, A. *et al.* Ballistic Transport in Graphene Antidot Lattices. *Nano Lett.* **15,** 8402–8406 (2015).

19. Yagi, R. *et al.* Ballistic transport in graphene antidot lattices. *Phys. Rev. B* **92,** 195406 (2015).

20. Zhang, Y., Kim, Y., Gilbert, M. J. & Mason, N. Electron transport in strain superlattices of graphene. *arXiv* **1703.05689,** (2017).

21. Yankowitz, M., Xue, J. & LeRoy, B. J. Graphene on hexagonal boron nitride. *J. Phys. Condens. Matter* **26,** 303201 (2014).

22. Lee, M. *et al.* Ballistic miniband conduction in a graphene superlattice. *Science* **353,** 1526–1529 (2016).

23. Wang, L. *et al.* One-Dimensional Electrical Contact to a Two-Dimensional Material. *Science* **342,** 614–617 (2013).

24. Scarabelli, D. *et al.* Fabrication of artificial graphene in a GaAs quantum heterostructure. *J. Vac. Sci. Technol. B* **33,** 06FG03 (2015).

25. Lee, G.-H. *et al.* Electron tunneling through atomically flat and ultrathin hexagonal boron nitride. *Appl. Phys. Lett.* **99,** 243114 (2011).

26. Britnell, L. *et al.* Electron tunneling through ultrathin boron nitride crystalline barriers. *Nano Lett.* **12,** 1707–1710 (2012).

27. Park, C. H., Yang, L., Son, Y. W., Cohen, M. L. & Louie, S. G. New generation of massless dirac fermions in graphene under external periodic potentials. *Phys. Rev. Lett.* **101,** 126804 (2008).

28. Hofstadter, D. R. Energy levels and wave functions of Bloch electrons in rational and irrational magnetic fields. *Phys. Rev. B* **14,** 2239–2249 (1976).

29. Claro, F. H. & Wannier, G. H. Magnetic subband structure of electrons in hexagonal lattices. *Phys.*

Rev. B **19,** 6068–6074 (1979).

30. Langbein, D. The Tight-Binding and the Nearly-Free-Electron Approach to Lattice Electrons in External Magnetic Fields. *Phys. Rev.* **180,** 633–648 (1969).

31. Wannier, G. H. A Result Not Dependent on Rationality for Bloch Electrons in a Magnetic Field. *Phys. Status Solidi* **88,** 757–765 (1978).

32. MacDonald, A. H. Landau-level subband structure of electrons on a square lattice. *Phys. Rev. B* **28,** 6713–6717 (1983).

33. Geim, A. K. & Grigorieva, I. V. Van der Waals heterostructures. *Nature* **499,** 419–425 (2014).